\documentclass[aps,12pt,showpacs]{revtex4}

\usepackage{graphicx}
\usepackage{dcolumn}
\usepackage{amssymb,amsmath}
\usepackage{natbib}
\usepackage{epstopdf}
\usepackage{color}

\def\be{\begin{equation}}
\def\ee{\end{equation}}
\def\bea{\begin{eqnarray}}
\def\eea{\end{eqnarray}}

\def\begineq{\begin{equation}}
\def\endeq{\end{equation}}

\begin{document}

\title{Velocity profiles in strongly turbulent Taylor-Couette flow}
\author{Siegfried Grossmann$^{1}$, 
Detlef Lohse$^{2}$, and 
 Chao Sun$^{2}$}
\affiliation{$^1$ Fachbereich Physik, Renthof 6, Philipps-Universitaet Marburg, D-35032 Marburg, Germany\\
$^{2}$ Department of Science \& Technology, Mesa+ Institute,  and J.\ M.\ Burgers Centre for Fluid Dynamics,\\ University of Twente, 7500 AE Enschede, The Netherlands}

\date{\today}

\begin{abstract}
We derive the velocity profiles in strongly turbulent Taylor-Couette flow for the general case of independently rotating cylinders. The theory is based on the Navier-Stokes equations in the appropriate (cylinder) geometry. In particular, we derive the 
 axial and the angular velocity profiles as functions of distance from the cylinder walls and find that both 
follow a logarithmic profile, with downwards-bending curvature corrections, which are more pronounced for the angular velocity profile
as compared to the axial velocity profile, and which strongly increase with decreasing ratio $\eta$ between inner and outer cylinder radius. 
In contrast, the azimuthal velocity does not follow a log-law. 
We then 
compare the angular and azimuthal velocity profiles  with the recently measured
 profiles in the ultimate state of (very) large Taylor numbers.  
Though the {\em qualitative} trends are the same -- down-bending for large wall distances and 
(properly shifted and non-dimensionalized) angular velocity profile $\omega^+(r)$
being closer to a log-law than (properly shifted and non-dimensionalized) azimuthal velocity profile $u^+_{\varphi}(r)$ -- 
{\em quantitative} 
 deviations are found for large wall distances.
 We  attribute these differences to the
Taylor rolls and the height dependence of the profiles, neither of which are considered in the theoretical approach.        
\end{abstract}

\pacs{47.27.-i, 47.27.te, 47.52.+j, 05.40.Jc}
\maketitle

\section{Introduction}
Having measured, analyzed, and discussed the {\em global} properties of the Rayleigh-B\'enard (RB) (cf.\ \cite{ahl09,loh10}) and of the Taylor-Couette (TC) \cite{gil11,hui12,gil12} devices, the two paradigmatic systems of fluid mechanics, which realize strongly turbulent laboratory flow, there is increasing interest in the local properties of these flows, e.\ g.\ in their flow profiles. In the ultimate state of RB thermal convection logarithmic profiles have been measured \cite{ahl12,fun09,ahl09b,he12,he12a} and calculated from the Navier-Stokes-equations \cite{gro12}. 

\begin{figure}[tb]
    \centering         
\includegraphics[scale=0.5]{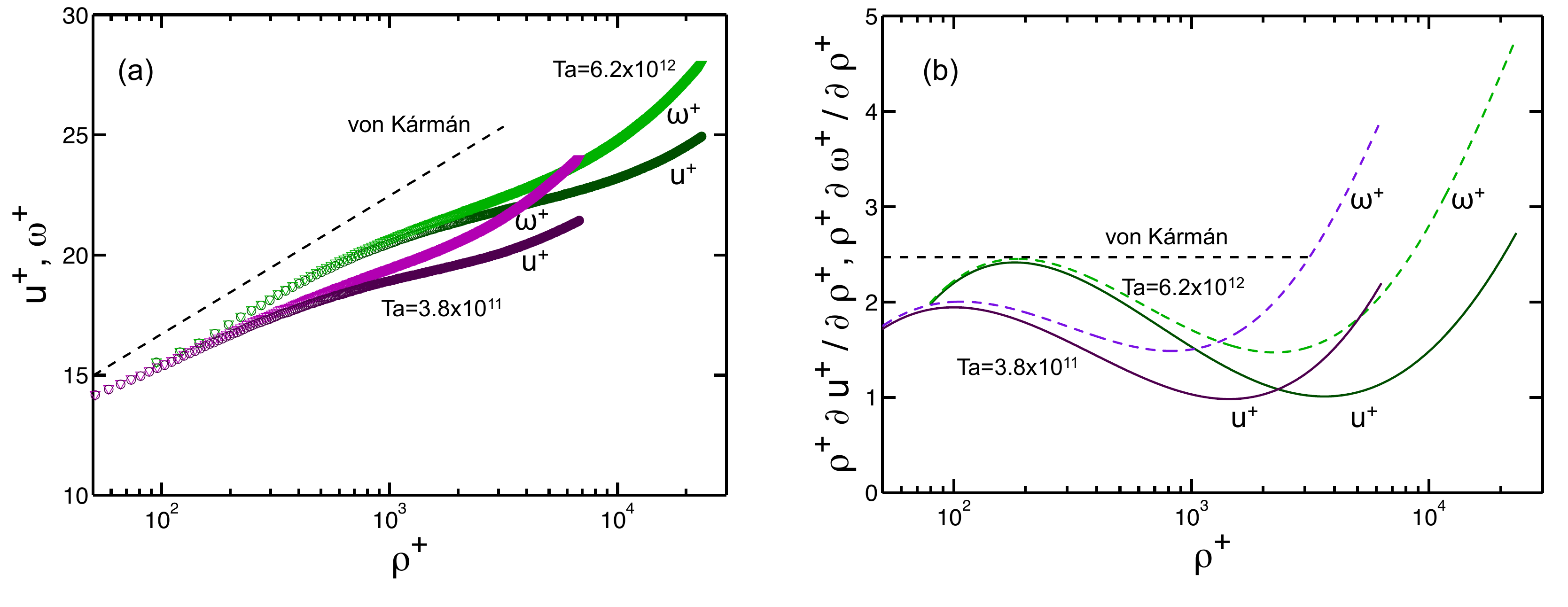}                    
    \caption{(color online)
(a) log-linear plot of the azimuthal velocity profiles in so-called wall units $u^+ (\rho^+)$ (see text) for two 
Taylor numbers $Ta = 6.2 \cdot 10^{12} $ and $Ta = 3.8 \cdot 10^{11}$ as measured in ref.\ \cite{hui13}, for fixed
height. 
In these units the profiles for various $Ta$ collapse  for small wall distances $\rho^+ $ in the viscous sublayer.   
To calculate the derivatives shown in (b) the data of ref.\ \cite{hui13} (or of (a)) have been fitted with a 5th order polynomial for smoothening. Also shown as a straight dashed line 
is the von K\'arm\'an log-law 
$ \kappa^{-1} \ln \rho^+ + B$, 
with the von K\'arm\'an constant $\kappa = 0.4$ and
the offset $B = 5.2$ \cite{hui13}. In addition, we show the two angular velocity profiles
$\omega^+ (\rho^+) $ resulting from the azimuthal velocity profiles, which obviously nearly overlap with $u^+ (\rho^+)$ for
small wall distances $\rho^+$, but for larger $\rho^+$ closer to the center range of the gap bend down less strongly. The data all extend to mid gap $d^+/2$.
(b) The compensated slopes  
$\rho^+ du^+/d\rho^+$ and 
$\rho^+ d\omega^+/d\rho^+$ of the azimuthal and angular velocity profiles, respectively, vs $\log_{10} \rho^+$ for the same 
curves as in figure (a). 
}
\label{fig1}
\end{figure}

In Taylor-Couette flow between independently rotating cylinders one can get considerably deeper into the ultimate range than in RB  flow, cf.\ \cite{gil12}, due to the better efficiency of mechanical driving as compared to thermal one. In this ultimate TC flow 
regime, i.\ e., for very large Taylor numbers $Ta \gtrsim 5\cdot 10^{8}$ \cite{ost13}, the profiles of the azimuthal velocity have recently been measured \cite{hui13} within the Twente Turbulent Taylor-Couette (T$^3$C) facility \cite{gil11a}. In figure \ref{fig1}a we 
reproduce the mean azimuthal velocity  profile at the inner cylinder for two large $Ta$ numbers. In ref.\ \cite{hui13} it was argued 
that the flow profiles roughly follow the von K\'arm\'an law \cite{pop00} for wall distances $\rho$ much larger than in the
viscous sublayer and much smaller than half of the gap, whose  width is  $d = r_o - r_i$. Indeed, as seen from figure \ref{fig1}a,  for the $Ta = 6.2 \cdot 10^{12}$
case and for $\log_{10} \rho^+  \approx  2.5$ -- two orders of magnitude smaller than the outer length scale which is the half width 
$d/2$ of the gap -- 
the azimuthal velocity profile $U_\varphi (r) $ after proper shifting seems to be possibly consistent with a log-law,
\begin{equation} u^+ (\rho^+) \equiv 
{\omega_i \cdot r_i \over u_i^*} - {U_\varphi (r) \over u_i^*} 
\approx \kappa^{-1} \ln \rho^+ + B,  \label{vonkarman} 
\end{equation}
over a small range, but for larger wall distances $\rho^+$ 
the curve  bends down towards smaller values. 
This behavior is pronouncedly  different from the standard 
pipe flow case \cite{pop00,mar10b,hul12,mar13,hul13b}, for which the profiles first bend up before they 
bend down towards the center of the flow. 
In equation (\ref{vonkarman}) the azimuthal velocity and the distance $\rho$ from the wall have been presented 
in the usual wall units $u_i^*$ and $\delta_i^*=\nu/  u_i^*$, marked with the usual superscript $^+$ and to be exactly defined later. 
In figure \ref{fig1}a
we also show the angular velocity profiles $\omega^+ (\rho^+)$ in the respective wall unit, resulting from the mean angular velocity $\Omega (r) = U_\varphi (r)/r$. 
Both the azimuthal velocity profile $u^+(\rho^+)$ as well as the angular velocity profile $\omega^+ (\rho^+) $
are shifted so that they are zero at the inner cylinder and then have positive slope. 
Also $\omega^+(\rho^+)$ is normalized with wall units, i.e., 
\be
\omega^+(\rho^+) \equiv 
{\omega_i - \Omega(\rho^+) \over \omega_i^*}
={\omega_i  r_i - {r_i\over r_i + \rho } u_\varphi \over u _i^*} 
= {1\over 1 + {\rho \over r_i} } u^+_\varphi(\rho^+)  + {\omega_i r_i \over u_i^*} { {\rho\over r_i}  \over 1 + {\rho\over r_i}}
, 
\label{logical-omega-plot} 
\ee
with $\omega_i^* = u_i^*/r_i$.  
In the regime
of the log-law $\omega^+(\rho^+)$ is nearly 
indistinguishable from the azimuthal velocity itself (see figure \ref{fig1}), since $\rho/r_i \ll 1$, but note  that 
obviously {\it not both $u^+(\rho^+)$ {\it and} $\omega^+(\rho^+)$} can follow a log-law,
due to the extra $\rho$-dependent factor in between them and the extra additive term. It is the last term in equation (2) which brings the $\omega^+$ profile above the $u^+$ profile with increasing $\rho/r_i$, because $\omega_ir_i/u_i^*$ will turn out to be significantly larger than one, e.g., it varies from 40 to 54 for Ta from 6 $\times10^{10}$ to 6 $\times 10^{12}$. 

One can calculate the difference between $\omega^+$ and $u^+$ by adding and subtracting properly in the first term of eq. (2) and finds 
\be \label{difference}
\omega^+(\rho^+) -  u^+(\rho^+)  =  \frac{U_{\varphi}}{u^*_{i}} \frac{\frac{\rho}{r_i}}{1 + \frac{\rho}{r_i}} ~.
\ee
The first factor will turn out to be (see table II, for $Ta = 6 \times 10^{12}$) between 54 near the wall and about 21.6 at midgap (see Figure 2 in Huisman {et al.} \cite{hui13}); the second factor varies between 0 at the cylinder and $(1-\eta)/(1+\eta)$, thus 0.166 for T$^3$C 
\cite{gil11a}
at midgap. 
For the difference (\ref{difference}) 
this gives an increase between 0 and about 3.6, which can be observed in Figure 1 (and also in Figure 2) and explains the increasing separation between $\omega^+$ and $u^+$. Note that $\omega^+$ is much nearer to the log-law than $u^+$ is.

The best way to test how well data follow a particular law is to introduce {\it compensated} plots, as has also been done
for structure functions \cite{gro97a} and for RB global scaling laws such as Nu or Re vs Rayleigh number Ra. Here, to test how well the data follow eq.\ (\ref{vonkarman}), rather than plotting
$u^+$ vs $\log_{10} \rho^+$ as done in fig.\ \ref{fig1}a, we plot the compensated slope $\rho^+ du^+/d\rho^+$ of the profile, 
see fig.\ \ref{fig1}b. If an exact log-law would hold exactly, this should be a constant horizontal line. From the 
figure we see that this
does not hold, neither for the azimuthal velocity $u^+$, nor for the angular velocity $\omega^+$. There is only a broader
maximum between $\log_{10} \rho^+ \approx 2.0$ and $2.3$ (depending on $Ta$), i.e., at a scale roughly two order of magnitude larger than the inner length scale and two orders of
magnitude smaller than the outer length scale.

Clearly, 
these data ask for a theoretical interpretation and explanation from the Navier-Stokes equations. 
For strongly driven RB flow such an explanation for the corresponding temperature profiles, which also show a logarithmic
profile \cite{ahl12}, 
 has already been offered previously in \cite{gro12}. In the present paper we shall derive the velocity profiles for strongly driven TC flow from the Navier-Stokes equations in the very same spirit and discuss their physics and features.  
In particular, we will check whether the experimentally observed down-bending of the azimuthal velocity profiles (and also of the angular velocity profiles) can be 
understood as a curvature effect, caused by the curvature of the wall, i.e., of the inner cylinder.  We will find that the profiles 
following from our theoretical approach indeed bend 
down, but weaker than experimentally found. Therefore the strong down-bending experimentally found in ref.\ \cite{hui13}
must have additional reasons. 

We start in the next section (II) by summarizing the Navier-Stokes based approach
 for the derivation of the profiles in cylinder coordinates. We then in section III shall derive and study the profile of the axial component $u_z$ versus wall distance $\rho = r - r_i$ or $\rho = r_o - r$. Here $r_{i,o}$ are the inner and outer cylinder radii, respectively. We analyze the axial component at first, since we consider this -- together with $u_r$ -- as the representative of the so-called ``wind'', responsible for the transport of the angular velocity $\omega = u_{\varphi}/r$, whose difference $\omega_i - \omega_o$ between the inner and outer cylinders drives the Taylor-Couette turbulence. Unfortunately, experimental data for the wind profile are not yet available, 
in contrast to the above mentioned measurements of the azimuthal component \cite{hui13}. Next, in section IV, the mean
azimuthal  velocity profile $U_{\varphi} (\rho) $ or rather the mean angular velocity profile $\Omega (\rho) = U_{\varphi}(\rho) /r$ versus $\rho$ is 
derived from the respective
 Navier-Stokes equation. Analogously to the temperature field 
in RB flow here in TC the angular velocity field $\omega$ is transported by the wind and by its fluctuations $u^*_i$ at the inner or $u^*_o$ at the outer cylinder, originating from the respective (kinetic) wall stress tensor element $\sigma_{rz}$. In section V we extend the 
 comparison with the  experimental data of ref.\ \cite{hui13} and then close with some concluding remarks in section VI.   

\section{Theoretical basis}

By detailed comparison of TC with RB flow we identify in this section the relevant quantities to calculate (and useful to measure). The theory has, of course, to be based on the Navier-Stokes equations for the three velocity components and the (kinetic) pressure field $p$ (equal to the physical pressure divided by the fixed density $\rho_{fluid}$ of the fluid). We repeat them here in the appropriate (cylinder) coordinates for the readers' convenience (cf.\ \cite{ll87}):   
\be \label{NS-phi}
\partial_t u_{\varphi} + (\vec{u} \cdot \vec{\nabla} ) u_{\varphi} + \frac{u_r u_{\varphi}}{r} = - \frac{1}{r} \partial_{\varphi} p + \nu \left( \Delta u_{\varphi} - \frac{u_{\varphi}}{r^2} + \frac{2}{r^2} \partial_{\varphi} u_r \right),
\ee
\be \label{NS-z}
\partial_t u_z +  (\vec{u} \cdot \vec{\nabla} ) u_z  = - \partial_z p + \nu \Delta u_z ~, 
\ee 
\be \label{NS-r}
\partial_t u_r +  (\vec{u} \cdot \vec{\nabla}) u_r - \frac{u_{\varphi}^2}{r} = - \partial_r p + \nu (\Delta u_r - 
\frac{u_{\varphi}}{r^2} - \frac{2}{r^2}\partial_{\varphi} u_{\varphi}) .
\ee
In addition, incompressibility is assumed. As usual the velocity fields are decomposed into their long-time means and their  fluctuations, whose correlations give rise to the Reynolds stresses, which will be modeled appropriately. We shall apply the well known mixing length ansatz \cite{ll87,pop00} and introduce turbulent viscosity and turbulent angular-momentum-diffusivity. All this then will lead to the respective profile equations. 
  
There are two basic differences between RB and TC flow. First, in contrast to RB, which is thermally driven by a temperature difference $\Delta = T_b - T_t$ between the bottom and top plate temperatures $T_{b,t}$, leading to a vertical temperature and a horizontal velocity (wind) profile, in TC flow there is a velocity (vector) field $\vec{u}(\vec{x},t)$ only. This is driven by a torque input due to different rotation frequencies $\omega_i$ and $\omega_o$ of the inner and outer cylinders.  Second, TC-flow 
can be compared to 
 non-Oberbeck-Boussinesq-(NOB)-flow (the more ``NOBness'', the smaller $\eta = r_i/r_o$), because its inner and outer boundary layers (BLs) have different profile slopes and thus BL-widths, as shown in \cite{eck07b} (which we henceforth cite as EGL). It is 
\be \label{slope-difference} 
r_i^3 \cdot \frac{\partial \omega}{\partial r} \Big|_i = r_o^3 \cdot \frac{\partial \omega}{\partial r} \Big|_o ~.
\ee
To take care of the different BL thicknesses we consider the inner and outer BLs separately. This does not require different physical parameters as for NOB effects in RB, for which the NOBness originates 
 from the temperature dependence of the
fluid properties, e.g.\ the 
 kinematic viscosity. In TC the kinematic viscosity $\nu$ is the same in both BLs; it is the boundary conditions which are different in TC, in particular 
the different wall curvatures at the inner and outer cylinders, leading to
different profile slopes.

The three velocity components in TC flow (instead of the velocity and temperature fields in RB flow) are subdivided into (a) the two components $u_r$ and $u_z$, known as the perpendicular components, and (b) the longitudinal component $u_{\varphi}$ or angular velocity $\omega = u_{\varphi} /r$. The former ones correspond to the convection or transport flow, the so-called ``wind'' field, the latter one to the thermal field in RB. This interpretation is based on the expression for the angular velocity current $J^{\omega}$ and the corresponding TC-Nusselt number $N^{\omega}$, which in TC play the role of the thermal current $J$ and Nusselt number $Nu$ in RB flow. In EGL \cite{eck07b} we have shown that
\be \label{current}
J^{\omega} = r^3 \left[ \langle u_r \omega \rangle_{A,t} - \nu \partial_r \langle \omega \rangle_{A,t}  \right] 
\ee
is r-independent and defines the (dimensionless) angular velocity current 
\be
N^{\omega} = J^{\omega} / J^{\omega}_{lam} . 
\ee
Here $J^{\omega}_{lam}$ denotes the analytically known angular momentum current in the laminar flow state of small Taylor number TC flow, see EGL \cite{eck07b}, eq.(3.11). The non-dimensional torque is 
\be \label{torque}
G = \nu^{-2} J^{\omega} = \nu^{-2} J^{\omega}_{lam} N^{\omega}, 
\ee  
which is related to the physical torque $\cal{T}$ by $\cal{T}$$= 2 \pi \ell \rho_{fluid} \nu^2 G = 2 \pi \ell \rho_{fluid} J^{\omega}$. The relation to the ($r,\varphi$)-component of the stress tensor is (e.g.\ at the inner cylinder)
\be \label{r-phi-stress}
\Pi_{r\varphi}(r_i) \equiv \rho_{fluid}~\sigma_{r\varphi}(r_i) = - \rho_{fluid} ~\nu ~r_i \left( \frac{\partial \omega}{\partial r} \right)_{r_i} = \rho_{fluid} r_i^{-2} J^{\omega} .
\ee
(For all this we refer to EGL \cite{eck07b}.) As TC flow is considered to be incompressible, we can 
always use the kinematic quantities and equations, i.e., after dividing by $\rho_{fluid}$, which then plays no explicit role anymore. In particular, we henceforth always use the kinetic stress tensor $\sigma_{ij}$.

The global transport properties depend on $\omega_i$ and $\omega_o$ in form of the Taylor number, which we define as 
\be \label{taylornumber}  
Ta = \frac{r_a^4}{r_g^4} ~ \frac{d^2 r_a^2 (\omega_i - \omega_o)^2}{\nu^2} .
\ee
Here $r_a = (r_i + r_o)/2$ is the arithmetic mean of the two cylinder radii and $r_g = \sqrt{r_i r_o}$ their geometric mean; $d = r_o - r_i$ is the gap width between the cylinders. 
\noindent In case of resting outer cylinder we in particular have
\be \label{taylor-reynolds}
Ta = \frac{r_a^6}{r_g^4 r_i^2} ~ Re_i^2 = \frac{\left( \frac{1+\eta}{2}\right)^6}{\eta^4} ~ Re_i^2 .
\ee
The inner cylinder Reynolds number is given by $Re_i = r_i \omega_i d / \nu$. Here 
$\eta$ is the radius ratio $\eta \equiv r_i/r_o \in (0,1)$ as usual. With respect to the coordinates we note the following correspondence between those for the top and bottom plates in RB samples as compared to the curved TC cylinder coordinates: It corresponds $x ~\mbox{in RB} \leftrightarrow z ~\mbox{in TC}$, stream wise direction; $z ~\mbox{in RB} \leftrightarrow  r ~\mbox{in TC}$, wall normal direction; and $y ~\mbox{in RB} \leftrightarrow  \varphi ~\mbox{in TC}$, lateral direction.    

While the angular velocity $\omega = u_{\varphi}/r$ in TC corresponds to the temperature field in RB, as already explained by
 EGL \cite{eck07b}, the transport flow or convection field, known as the wind, is described by the components $u_r$ and
$u_{\varphi}$. The Taylor roles (or their remnants in the turbulent state) correspond to the RB-rolls. The wind $u_x (z)$ in RB has a profile as a function of height z, while the component $u_z$ in TC has a profile as a function of r or rather of $\rho = r - r_i$ or $\rho = r_o - r$, which measure the wall normal distance. The up and down flow along the side walls of RB has its analogue in the $u_r$ component of TC flow. In contrast to the mainly studied aspect ratio $\Gamma = 1$ samples (or $\Gamma$ of order 1) in RB, in TC we usually have larger $\Gamma$  (order 10 or more). Thus there are more than only one Taylor role remnants in TC. We shall have in mind one of those as 
representative. With all these identifications we 
shall decompose the flow field components into their long(er)-time means and their fluctuations as follows: $u_{\varphi} = U_{\varphi}(r) + u'_{\varphi} = r \Omega(r) + u'_{\varphi}$, $u_z = U_z(r) + u'_z$, and $u_r = u'_r$, where the fluctuations $u'$ still depend on the full coordinates $\vec{x}$ and $t$. There is a mean angular velocity flow profile $\Omega(r)$, there is also a mean axial flow profile $U_z(r)$ at least within each roll remnant, but there is no longer-time mean radial flow component across the gap in a roll remnant.

\section{The wind profile}

Using the correspondences just described we have to study the axial component's time-mean $U_z$ 
 as a function of inner cylinder wall distance $\rho = r-r_i$ in order to derive and understand the profile of the wind field near the inner cylinder. Time averaging the $z$-equation (\ref{NS-z}) we have $\partial_t \hat{=} 0$, $\partial_{\varphi} \hat{=} 0$, and in the assumed approximation no height dependence $\partial_z \hat{=} 0$. There also is no axial pressure drop, i.e., $\partial_z p = 0$. With all this the viscous 
term of (\ref{NS-z}) is $ \nu \frac{1}{r} \partial_r r \partial_r U_z(r)$. The nonlinear terms (with the 
continuity equation) can be rewritten as $ \overline{(\vec{u} \cdot \vec{\nabla} ) u_z}^t  = \overline{\vec{\nabla}\cdot \vec{u} u_z}^t =\frac{1}{r} \partial_r (r \overline{u_r' u_z'}^t)$. Putting both contributions together, the Navier-Stokes equation for the wind profile reduces to $\frac{1}{r} \partial_r [...] = 0$ or $[...] \equiv r \overline{u'_r u'_z}^t 
- \nu r \partial_r U_z =$ constant. As there are no Reynolds stress contributions at the cylinder walls, we find
\be \label{wind-fluc-scale}  
\nu r_o \partial_r U_z(r_o) = \nu r_i \partial_r U_z(r_i) \equiv r_i (u^*_{z,i})^2 = r_o (u^*_{z,o})^2 ,   
\ee
which defines the wind fluctuation scales $u^*_{(z,i),(z,o)}$ in terms of the inner and outer cylinder kinetic wall stress tensor component $\sigma_{rz}(r_{i,o}) \equiv \nu \partial_r U_z(r_{i,o})$ (cf.\ \cite{ll87}, Sect.\ 16). Note that from eq.\ 
(\ref{wind-fluc-scale}) it follows that the wind fluctuation amplitudes are {\em different} at the two cylinders: $u^*_{z,o} / u^*_{z,i} = \sqrt{r_i / r_o} = \sqrt{\eta} \neq 1$ in the TC system. Depending on the radius ratio $\eta$ the wind fluctuation amplitude is thus somewhat weaker in the outer cylinder boundary layer (BL) than in the inner one. One may interpret this as more space being available.

While the velocity fluctuation amplitudes $u^*_{(z,i),(z,o)}$ are defined in terms of the $rz$-wall stress, independent of the Reynolds stress, this latter one acts in the interior of the flow. Thus for determining the wind profile an ansatz is needed for it. The Reynolds stress is, of course,  responsible for the turbulent viscosity in the convective transport. We assume that the mixing length idea can be used for TC flow, too, and write
\be \label{reynoldsstress}
\overline{u_z' u_r'} = - \nu_{turb}(r) \partial_r U_z . 
\ee      
We furthermore assume the validity of the mixing length modeling for the turbulent viscosity $\nu_{turb}(r)$, considering it as depending on the wall distance as the characteristic length scale and the velocity fluctuation amplitude as the characteristic velocity 
scale, 
\be \label{turb-viscosity}
\nu_{turb} (r) ~ = ~K^T_i  (r - r_i)  u^*_{z,i}  ~~~\mbox{and}~~~  = ~K^T_o  (r_o - r)  u^*_{z,o} ,     
\ee 
respectively. Here $K^T_{i,o}$ are non-dimensional constants, denoted as transversal von K\'arm\'an constants, possibly different for the inner and outer cylinders. 

Let us now, for simplicity, concentrate on the inner cylinder; the respective outer cylinder formulas are straightforward then. With the said ansatz the wind profile is determined by the equation $(\nu + \nu_{turb}(r)) \cdot r\partial_r U_z = r_i \cdot (u^*_{z,i})^2$ or 
\be  \label{wind-profile-r}
r ~\partial_r ~U_z(r) = \frac{r_i \cdot(u^*_{z,i})^2}{\nu + K^T_i \cdot (r-r_i) \cdot u^*_{z,i}} ~.
\ee            
The relevant length scale is the distance $\rho=r-r_i \ge 0$ from the cylinder wall, i. e., $r = r_i + \rho$. Defining the characteristic viscous wall distance(s) 
\be \label{viscous-scale}
\delta^*_{(z,i),(z,o)} \equiv \frac{\nu}{u^*_{(z,i),(z,o)}}  
\ee 
at which $\nu_{turb}(r)$ is of the order of the molecular viscosity $\nu$, we can introduce wall units as usual, 
\be \label{wallunits}
\rho^+_{(z,i),(z,o)} \equiv \rho/\delta^*_{(z,i),(z,o)} ~~~\mbox{and}~~~U^+_{(z,i),(z,o)} \equiv U_z(\rho)/ u^*_{(z,i),(z,o)}.
\ee 
Then the profile slope equation(s) for the wind in axial direction near the cylinder wall(s) as a function of the respective wall distance in wall units reads 
\be \label{windprofileequation}
\frac{d U^+}{d \rho^+}~=~\frac{1}{(1+\rho^+/r^+_{i,o})(1+ K^T_{i,o} \rho^+)}.
\ee 
The first factor in the denominator is the factor r from the lhs of eq.\ (\ref{wind-profile-r}) and  $r^+_{i,o}$ denotes the inner (or outer) cylinder radius in the respective wall units. Usually the characteristic wall distance is rather small, 
 $\rho^+/ r^+_{i,o} \ll 1$, of course unless the inner cylinder is very thin, $\eta \approx 0$.  

Let us now draw conclusions:

(i) For sufficiently small distances $\rho^+ \ll 1$ {\em and} $\rho^+ \ll r^+_{i,o}$ we find the viscous, linear sublayer as usual, 
\be \label{viscous-sublayer}
U^+ = \rho^+  = \rho/ \delta^*_z.
\ee  
If it were possible to 
measure the slopes of the viscous, linear sublayers, one would be able to immediately
 determine the viscous length scales and therefore also  the velocity fluctuation scales $u^*_{(z,i),(z,o)} = \nu / \delta^*_{(z,i),(z,o)}$.  

(ii) In general, we can decompose the fraction in eq.\ (\ref{windprofileequation}) into partial fractions and find the profile as a sum of two log-terms,  
\be \label{windprofile}
U^+(\rho^+) = \frac{1}{K^T} \cdot \frac{\mbox{ln}(1+ K^T \rho^+) - \mbox{ln}(1+\rho^+/r^+_{i,o})}{1 - 1/(r^+_{i,o} K^T)} .
\ee
Also here $K^T$ means $K^T_{i,o}$. This solution for the wind profile satisfies the boundary condition at the cylinder wall $U^+(0) = 0$ and also reproduces the linear viscous sublayer law $U^+ = \rho^+$ for small $\rho^+$. -- The case $r^+_{i,o} = 1/K^T_{i,o}$ deserves special care, see (iii).

(iii) If one of the relations either for the inner or for the outer cylinder is valid, $r^+_{i,o} = 1/K^T_{i,o}$ or $r_{(z,i),(z,o)} = \delta^*_{i,0} /K^T = \nu / (u^*_{(z,i),(z,o)}~K^T)$, i. e., for tiny inner or outer cylinder radius $r_{i,o}$, the profile slope is $dU^+ / d\rho^+ = 1/(1+ K^T \rho^+)^2]$. Therefore 
$U^+(\rho^+) = \rho^+ / (1+K^T \rho^+)$
and for large $\rho^+ \gg 1$ 
there is {\em no} log-profile in this special case but instead $U(\rho^+) $ is $\rho^+$-independent. 
This case is obviously more a mathematical pecularity, rather than being physically relevant.

(iv) The main difference between the wind profile in curved TC flow and that of plane plate flow (e.g.\ in RB) is the factor of $r$
 in the profile equation (\ref{wind-profile-r}) or $(1 + \rho^+/r^+_{i,o})$ in the profile equation (\ref{windprofileequation}). Now, $r = r_i + \rho$ varies between $r_i$ and $r_i + d/2$; beyond, for even larger $\rho$, one is in the outer part of the gap. Therefore the relative deviation of $r$ from $r_i$ is at most $d/(2r_i)$ or $0.5(\eta^{-1}-1)$. If this is less than the
experimental precision of say 20\%; 10\%; 5\%, 
 the curvature is unobservable. This happens for all  radius ratios $\eta$ less than some characteristic, precision dependent value $\eta_e > 0.714; 0.833; 0.909$. The smaller the observable relative deviation is, the larger the characteristic $\eta_e$ or the smaller the characteristic gap width must be. For  $\eta > \eta_e$ up to $\approx$ 1 the experimental uncertainty hides the curvature effects in the wind profile. -- This estimate
must even be sharpened for the observation of deviations from the log-layer, since this does not extend until gap half width, thus increasing the requirements for experimental identification of the curvature effects in the log-layer range.   

(v) In general, $r^+_{i,o}$ will be large since $r_{i,o} \gg \delta^*_{(z,i),(z,o)}$. We shall confirm this below with an estimate of $u^*_{(z,i),(z,o)}$. Then the implication of the finite curvature radius $r^+_{i,o}$ of the cylinder walls can be discussed as follows: The factor $1 + \rho^+/r^+_{i,o}$ in the denominator of the profile equation (\ref{windprofileequation}) varies between 1 and $1+d^+/(2r^+_{i,o}) = (1+\eta^{-1})/2$. (Analogously for the outer cylinder $1 + d^+/(2r^+_o) = 1 + d/(2r_o) = (3-\eta)/2$~). For the Twente $T^3C$ facility with its radius ratio $\eta = 0.7158$ the factor $1+\rho^+/r^+_{i,o}$ for the inner BL varies between 1 and $1.199 \approx 1.20$. 
For the outer cylinder the corresponding slope modification factor is 1.14. As expected the curvature effects are always smaller at the outer than at the inner cylinder. -- The profile equation thus describes a log-layer slope modified by a slightly decreasing (or an increasingly smaller) slope.      

The slope decrease will be the stronger the smaller the radius ratio $\eta$ is. In order to have $d/(2r_i) = 1$ (or even 5) one needs to consider $\eta = 1/3 = 0.333$ (or even $\eta = 1/11 = 0.091$). The smaller $\eta$, the better the curvature effects at the inner cylinder are visible. In contrast, for $\eta \rightarrow 1$, plane channel flow, there is no slope decrease anymore; there is then the {\it pure} log-law of the wall for the wind profile. 

We close this section on the wind profile by estimating the fluctuation amplitude(s) $u^*_{(z,i),(z,o)}$ and thus the viscous scales. To be specific we again consider the $T^3C$ facility \cite{gil11a}. Its working fluid is water with $\nu = 10^{-6} m^2s^{-1}$. Its geometric parameters are $r_i = 0.2000$m, $r_o = 0.2794$m, $d = r_o - r_i = 0.0794$m, its radius ratio is $\eta  \approx 0.7158$. In order to estimate the size of the fluctuation amplitude $u^*_{z,i}$ we write this as $u^*_{z,i} = \frac{u^*_{z,i}}{U_i}\cdot Re_i \cdot \frac{\nu}{d}$ with the inner cylinder velocity $U_i$ and the corresponding Reynolds number $Re_i = \frac{U_i d}{\nu}$. The outer cylinder be at rest, i.e., $Re_i$ characterizes the flow stirring. 

Now we have to estimate the ratio $u^*_{z,i}/U_i$ for various $Ta$ or $Re_i$, respectively. In \cite{gro11} we have derived an explicit expression for $u^*_z/U$ as function of $Re$ (and have applied it to RB flow in \cite{gro11,gro12}). 
In lack of any measurements for $u^*_z/U_i$ for the wind fluctuation scale in TC flow, we have to build on those RB  estimates. 
Since the fluctuation velocity $u^*_z$ is determined by the wall stress, only the immediate neighborhood of the cylinders is felt by the flow field, i.e., we might neglect the curvature and calculate $u^*$ as for plane flow. According to \cite{gro11} the relative fluctuation strength then is given by 
\be \label{lambert} 
\frac{u^*_z}{U} = \frac{\bar{\kappa}}{W(\frac{\bar{\kappa}}{b}Re)} ~~~\mbox{with}~~~ b \equiv e^{- \bar{\kappa} B}.
\ee
$B$ is the logarithmic intercept of the common log-law of the wall $u^+ = \frac{1}{\bar{\kappa}} \mbox{ln} z^+ + B$. We use $\bar{\kappa} = 0.4$ and $B = 5.2$ (cf.\ \cite{ll87,pop00}) which implies $b = e^{- \bar{\kappa} B} = 0.125$. The argument of Lambert's function $W$ then is $3.2 \times Re$. Depending on the values of the constants $\bar{\kappa}$ and $b$, which are taken here from pipe flow, channel flow, or flow along plates but have not yet been measured for TC flow, the fluctuation amplitudes $u^*_{z,i}/U_i$ at the inner (or outer) cylinder wall might differ slightly.

 Our results are compiled in table \ref{table1} for various $Ta$ and the respective $Re_i$ in the first two columns. Since in the present case of resting outer cylinder the relation between $Ta$ and $Re_i$ is given by eq.\ (\ref{taylor-reynolds}), for the $T^3C$ facility with above $\eta$ we in particular have $Ta = 1.5186~Re_i^2$. Column 3 offers ${u^*_{z,i}}/{U_i}$. This allows us
to determine the corresponding 
$u^*_{z,i}$ 
 shown in column 4. From that we obtain the respective viscous length scales $\delta^*_{z,i} = \nu/u^*_{z,i}$ (see column 5). Knowing all this we can determine also $r^+_i = r_i/\delta^*_{z,i}$ and $d^+/2$ and $d^+/100$ (the wall distance where  experimentally 
an approximate log-law for the azimuthal velocity $u_\varphi (\rho)$ had been found in ref.\ \cite{hui13}, see next section), 
all compiled in columns 6, 7, and 8  of table \ref{table1}. 

 \begin{table}[tb]
 \begin{center}
 \begin{tabular}{|c|c|c|c|c|c|c|c|}
 \hline
          $Ta$
       &  $Re_i$ 
       &  $\frac{u^*_{z,i}}{U_i}$
       &  $u^*_{z,i}$ in ms$^{-1}$
       &  $\delta^*_{z,i} = \frac{\nu}{u^*_{z,i}}$ 
       &  $\frac{r_i}{\delta^*_{z,i}}$
       &  $\frac{d/2}{\delta^*_{z,i}}$
       &  $\frac{d/100}{\delta^*_{z,i}}$
\\
\hline  
%          $4.6 \times 10^7$
%       &  $5.5 \times 10^3$ 
%       &  0.05174
%       &  0.003584 
%       &  $279 \times 10^{-6}$ m
%       &  717
%       &  142
%       & 3
%\\
%          $1 \times 10^9$
%       &  $2.6 \times 10^4$ 
%       &  0.04387
%       &  0.01437
%       &  $69.6 \times 10^{-6}$ m 
%       &  2874
%       &  570   
%       &  11
%\\
          $6 \times 10^{10}$  
       &  $2.0 \times 10^5$ 
       &  0.03645
       &  0.09181  
	   &  $10.9 \times 10^{-6}$ m
	   &  18350
	   &  3642
           &  73
\\
          $4.6 \times 10^{11}$
       &  $5.5 \times 10^5$
       &  0.03360
       &  0.23275
       &  $4.30 \times 10^{-6}$ m
       &  46510
       &  9233
       &  185
\\  
          $3 \times 10^{12}$
       &  $1.4 \times 10^6 $
       &  0.03133
       &  0.55242
       &  $1.81 \times 10^{-6}$ m
       &  110500
       &  21930
       &  439
\\
          $6 \times 10^{12}$
       &  $2.0 \times 10^6$
       &  0.03054
       &  0.76927
       &  $1.30 \times 10^{-6}$ m
       &  153850 
       &  30540
       &  611
\\
\hline
 \end{tabular}
 \end{center}
 \caption{Values for the wind fluctuation amplitude $u^*_{z,i}$ for some Taylor numbers, the corresponding $Re_i = 0.8115 \sqrt{Ta}$, also the respective viscous length scales and the inner cylinder radii and the gap half widths in wall units. For details see text. Note that $d^+/(2 r^+_i) = d/(2r_i) = 0.1985$. The values for $u^*_{z,i} / U_i$ have been calculated with eq.\ (\ref{lambert}), cf.\ \cite{gro11}.
}
\label{table1} 
\end{table}

A final remark concerning the relevant Reynolds number $Re$ for calculating $u_{z,i}^*$. One might argue that instead of the inner cylinder Reynolds number $Re_i$ one better should use the so-called wind Reynolds number $Re_w$, introduced in reference \cite{gil12}, page 130. For resting outer cylinder, i.e., for $\mu  \equiv \omega_o / \omega_i = 0$, this is $Re_w = 0.0424 \cdot Ta^{0.495}$ for the $T^3C$-geometry. This is roughly 5\% of $Re_i$. Since $Re_w$ is
 significantly smaller than the inner cylinder Reynolds number $Re_i$, one needs much larger $Ta$ to realize the Reynolds numbers in column 2 of table \ref{table1}. Also there is a significant difference between the RB-wind and the TC-wind: While in RB the wind is the only coherent fluid motion available in the (otherwise resting) system, in TC there is an intrinsic stimulus for fluid motion due to the 
inner cylinder rotation (or in general the difference in the rotation frequencies of the two cylinders). Thus there are two different velocities available, the wind $U_w$ {\it and} the inner cylinder velocity $U_i$. To improve insight, Table \ref{table4} in the appendix provides detailed numbers for the fluctuation amplitude due to the wind $U_w$ instead of the inner cylinder velocity $U_i$.     

In any case, presently no experimental data on the wind velocity profiles are available for TC flow. So we do not know whether 
the predicted log-profile with curvature corrections (\ref{windprofile}) 
exists and, if so, how far it will extend towards the gap center. To detect the curvature corrections experimentally,
a far extension towards the center will be crucial (as otherwise the correction factor will be too close to 1), 
and, as discussed above, obviously a small value of $\eta$ -- 
strong geometric NOBness -- will help to. In the next section 
we will discuss these issues in much more detail for the angular velocity profile,
for which experimental data exist.

\section{The angular velocity profile}

In TC flow, as has been explained, the mean angular velocity profile $\Omega(r)$ -- and not the azimuthal velocity $U_{\varphi}(r)$ -- corresponds to the temperature profile in RB thermal convection. This conclusion, as has been detailed in section II, is based on the comparison of the respective expressions for the transport currents, 
which one can derive from the Navier-Stokes and Boussinesq equations. To calculate the $\Omega$-profile in TC flow we start from the equation of motion (\ref{NS-phi}) for the azimuthal velocity $u_{\varphi}(\vec{x},t)$. Again we decompose the equation
into the long(er)-time mean and the fluctuations, $u_{\varphi} = U_{\varphi} + u'_{\varphi} = r\Omega(r) + u'_{\varphi}$. Again we have $\partial_t \hat{=} 0, \partial_{\varphi} p = 0, \partial_{\varphi} U_r =0$ and arrive at
\be \label{timeaveragedazimuthal}
\overline{(\vec{u}\cdot \vec{\nabla}) u_{\varphi}}^t + \frac{\overline{u_r u_{\varphi}}^t}{r} = \nu \left(\Delta U_{\varphi} - \frac{U_{\varphi}}{r^2}\right).
\ee
Reorganize the nonlinear lhs: $\overline{(\vec{u}\cdot\vec{\nabla})u_{\varphi}}^t + \frac{\overline{u_r u_{\varphi}}^t}{r} = \vec{\nabla}\cdot \overline{(\vec{u} u_{\varphi})}^t + \frac{\overline{u_r u_{\varphi}}^t}{r} = \frac{1}{r} \partial_r (r \overline{u_r u_{\varphi}}^t) + \frac{\overline{u_r u_{\varphi}}^t}{r} = \partial_r \overline{u_r u_{\varphi}}^t + 2 \frac{\overline{u_r u_{\varphi}}^t}{r} = \frac{1}{r^2} \partial_r (r^2 \overline{u_r u_{\varphi}}^t)$. Then reorganize the rhs: $\nu (\Delta U_{\varphi} - \frac{U_{\varphi}}{r^2}) = \nu (\frac{1}{r} \partial_r r \partial_r U_{\varphi} - \frac{U_{\varphi}}{r^2}) = \frac{\nu}{r^2}(r \partial_r r \partial_r U_{\varphi} - U_{\varphi}) = \frac{\nu}{r^2} \partial_r (r^3 \partial_r \frac{U_{\varphi}}{r})$. Thus time averaging leads to $r^2 \overline{u'_r u_{\varphi}'}^t  - \nu r^3 \partial_r \frac{U_{\varphi}}{r}$ = constant. 
A very similar expression is well-known from the derivation of the angular velocity current, see
EGL \cite{eck07b}. Apparently it is $\frac{U_{\varphi}}{r}$, i.\ e.\ $\Omega(r)$, the angular velocity, which is relevant, since it is $\partial_r \Omega$ and not $\partial_r U_{\varphi}$ which determines the current as well as the profile(s) near the wall(s), as just derived. 

Also the nonlinear term can be expressed in terms of $\omega'$, by separating a factor of $r$ from $u_{\varphi}'$. For the corresponding Reynolds stress we suggest the ansatz $\overline{u'_r \omega'}^t = - \kappa_{turb}(r) \partial_r \Omega(r)$. The turbulent transport coefficient $\kappa_{turb}(r)$ has dimension $\ell^2 / t$; we call it the turbulent $\omega$-diffusivity (in analogy to the turbulent 
temperature diffusivity). Having thus modeled the $\omega$-Reynolds stress, the $\Omega$ profile satisfies the equation
\be \label{omegaequation-1}
r^3 \left( \nu + \kappa_{turb}(r) \right) \partial_r \Omega(r)  = r^3_{i,o} \nu \partial_r \Omega \big|_{i,o} = - J^\omega .   
\ee
(The Reynolds stress $\overline{u' \omega'}^t$ does not contribute at the cylinder walls $r_{i,o}$.) 
This results in the profile equation
\be \label{omega-eq}
\partial_r \Omega(r) = \frac{- J^{\omega}}{r^3(\nu + \kappa_{turb}(r))} . 
\ee
If $J^{\omega}$ is positive, i. e., transport from the inner to the outer cylinder, the $\Omega$-profile decreases with $r$, as it should be. 

We now have to model the turbulent $\omega$-diffusivity $\kappa_{turb}(r)$. It seems reasonable to again use the mixing length ansatz, saying that $\kappa_{turb} (\rho ) \propto$ distance $\rho = r - r_i$ from the wall times a characteristic fluctuation velocity. 
But this time (i.e., for the angular velocity rather than for the wind velocity) 
 there are {\it two} candidates for such a characteristic velocity amplitude. First, again there is $u^*_{(z,i),(z,o)}$, the transverse velocity fluctuation amplitude due to the (kinetic) wall stress tensor component $\sigma_{rz}(r_{i,0})$, responsible for the wind profile $U_z(r)$ as discussed in the previous section III, where $(u_z^*)^2 = \sigma_{rz} = \nu \partial_r U_z$. 
However now, in addition, there is another wall stress induced fluctuation amplitude, because $J^{\omega} / r^2_{i,o} = r_{i,o} \nu \partial_r \omega$ also has the dimension of a squared velocity. Note that $J^{\omega} / r_{i,o}^2 = \sigma_{r\varphi}(r_{i,o}) \equiv (u^*_{i,o} )^2$ is the $r,\varphi$-component of the (kinetic) wall stress tensor, cf.\  \cite{ll87}, Section 15, eq.\ (15.17) and also EGL \cite{eck07b}, Section 3, eq.\ (3.5). We address $u^*_{i,o}$ as the longitudinal velocity fluctuation amplitude. 

It deserves experimental check or theoretical proof, whether the longitudinal velocity fluctuation $u^*$ and the transversal one $u^*_z$ are of equal size or are different. An argument for the former is that the Navier-Stokes equations couple all velocity components so strongly that they all fluctuate with the same amplitude. Another one in the same direction is that both
 $\sigma_{rz}$ and $\sigma_{r\varphi}$ express the (kinetic) shear along the cylinder wall, one in axial (stream-wise), the other one in azimuthal (lateral) direction. 

Thus there are two possible expressions for the turbulent $\omega$-diffusivity: $\kappa_{turb} \propto \rho \cdot u^*_z$ or $\kappa_{turb} \propto \rho \cdot u^*$. Until sufficiently clarified we use the latter one, being aware that the remaining constants just differ by the factor of $u^*_z/u^*$, possibly depending on the Taylor number $Ta$. We shall find, see Table \ref{table2}, that this ratio (for the inner cylinder) turns out to slightly increase with $Ta$, in the given $Ta$ range increasing from 1.3 to 1.6. That its
 deviation  from 1 might have its origin in an insufficient estimate of $u^*_z$, in which the parameters b and $\bar{\kappa}$ had to be guessed from pipe, channel, or plate flow, cf.\ Sec.\ III. If these fit parameters would depend on $Ta$, that would be reflected in a $Ta$-dependence of the fluctuation amplitude ratio $u^*_{z,i}/u^*_{i}$.   

Our ansatz for the turbulent $\omega$-diffusivity thus is
\be \label{omega-diffusivity}
\kappa_{turb} (\rho ) = K^L \rho ~u^*  ~~~\mbox{with}~~~ \rho = r - r_i ~ ~~\mbox{or} ~~~ \rho = r_o - r .
\ee 
The longitudinal von K\'arm\'an constant $K^L$ may or may not depend on Taylor number, as does the transversal von K\'arm\'an constant $K^T$ from eq.\ (\ref{turb-viscosity}). 
 Again, $u^*$ and $K^L$ may be different for the inner and outer boundary layers. In the following, for simplicity, we have in mind the inner cylinder, omitting the label i, but the corresponding equations hold for the outer BL, respectively. 

Introduce now as usual the viscous length scale 
\be \label{viscous-scales}
\delta^* = {\nu}/{u^*} .
\ee
The wall distance and the inner cylinder radius in $\omega$-wall units are 
\be \label{wallunitdistance}
\rho^+ = \rho / \delta^*;  ~~ r^+_i = r_i / \delta^* ~.
\ee
As a normalized angular velocity which increases with distance from the wall we define   
\be \label{wallunitomega}
\tilde \Omega := \frac{\omega_i - \Omega(r)}{\omega_i} .
\ee
This dimensionless profile $\tilde\Omega$ is zero at the cylinder surface and increases with increasing wall distance $\rho^+$. In contrast to $\omega^+$ it is normalized with the inner cylinder rotation frequency $\omega_i$ instead of $u^*_i / r_i \equiv \omega^*_i$. The equation for $\tilde \Omega(\rho^+)$, from eq.(\ref{omega-eq}), then reads 
\be \label{walluniteq}
\frac{d \tilde \Omega}{d (\rho^+/r_i^+)} = F_i ~\frac{1}{(\frac{1}{r_i^+} + K^L \frac{\rho^+}{r_i^+} ) \left( 1 + \frac{\rho^+}{r^+_i} \right)^3 } ~, \ee
where the distance from the wall now has been expressed in terms of 
\be
x= \rho^+/r_i^+ = \rho / r_i . \label{x} 
\ee
Here the dimensionless constant $F_i$, the slope factor of the profile equation, is defined as 
\be \label{wall-omega-nusselt}
F_i \equiv \frac{J^{\omega} \delta^*_{i}}{r_i^3 \omega_i \nu} = \frac{2(1-\mu )}{1-\eta^2} \cdot \frac{N^{\omega}}{r^+_i} .
\ee
Up to geometric features ($\eta$ and $r^+_i$) and the rotation ratio $\mu = \omega_o / \omega_i$, $F_i$ is just the $\omega$-Nusselt number $N^{\omega}$. To derive the slope factor $F_i$ from eq.\
(\ref{omega-eq}) one writes $J^{\omega}$ as $J^{\omega}_{lam} \cdot N^{\omega}$ and 
 then uses $J^{\omega}_{lam} = 2 \nu r_i^2 r_o^2 \frac{\omega_i - \omega_o}{r_o^2 - r_i^2}$ from eq.\ (3.11) in EGL \cite{eck07b}.
Note that the slope factor $F_i$ does not depend on the longitudinal von K\'arm\'an constant $K^L$.

Yet another form of $F_i$ is of interest. Reminding $J^{\omega} /r^2_i = (u^*_i)^2$ and using the definition 
 $\delta^*_{i} = {\nu}/{u^*_i}$ of the inner length scale, one arrives at 
\be \label{wall-omega-nusselt-2}
F_i = \frac{u^*_i}{r_i \omega_i} = \frac{u^*_i}{U_i} = \frac{\omega^*_i}{\omega_i} .  
\ee
Here we have introduced the fluctuation scale $\omega^*_i$ of the angular velocity by 
\be \label{omegafluctuationscale}
\omega^*_i \equiv {u^*_i}/{r_i} .
\ee
It is the very ratio of the angular velocity fluctuation amplitude $\omega^*_i$ and the inner cylinder rotation rate $\omega_i$ which measures the size $F_i$ of the $\tilde\Omega$-profile slope. Next, $F_i$ can be incorporated in the normalization of the profile,  giving the profile in the usual wall units, 
\be \label{smallomegawall}
\omega^+ (\rho^+ ) \equiv \frac{\omega_i - \Omega(r)}{\omega^*_i} ,
\ee
as already anticipated in equation (\ref{logical-omega-plot}).  
$\omega^+ (\rho^+ )$ satisfies the  profile equation in wall units, 
\be 
\label{profileequationwallunits}
\frac{d \omega^+ (\rho^+)}{d \rho^+} = \frac{1}{(1 + K^L \rho^+) \left( 1 + \frac{\rho^+}{r^+_i} \right)^3 }.
\ee
Expressed in terms of $x= \rho^+/r_i^+ = \rho / r_i$ this equation reads
\be 
\label{profileequationwallunits-prime}
\frac{d \omega^+ (x) }{d x} = \frac{1}{(\frac{1}{r_i^+} + K^L x ) \left( 1 + x \right)^3 } .
\ee
The analogous formulae hold for the outer cylinder. 
The advantage of this latter representation (\ref{profileequationwallunits-prime})
in terms of $x$ rather than in terms of $\rho^+$ is that -- apart from the very small correction $1/r_i^+$ -- the profile  
(\ref{profileequationwallunits-prime}) is universal, i.e., valid for all $Ta$. 
Such universal measure $x$ for the wall distance (cf.\ eq. (\ref{x})) can be introduced in TC Ð in contrast to the plate flow case -- as $r_i$ (or $r_o$) serve as a natural length unit, presenting the curvature radii of the walls.

We now  discuss  the obtained results on the slope of the angular velocity: 

(i) If both conditions  $\rho^+ \ll r^+_{i,o}$ and $\rho^+ \ll 1/K^L_{i,o}$ hold, one finds, as expected, the
 linear, viscous sublayer also for the angular momentum profile, $\omega^+ = \rho^+$. 

(ii) In case of rotation ratio $\mu  = 1$, i.e., $\omega_o = \omega_i$, the slope is $F_i = 0$ and from eq.\ (\ref{walluniteq}) we obtain $\frac{d \tilde \Omega}{d \rho^+} = \frac{d \omega^+}{d \rho^+} = 0$. The angular velocity thus is constant, we have solid body rotation.

(iii) As in general $r^+_{i,o}$ -- the inner (or outer) cylinder radius in terms of the tiny viscous scales -- is large, $1 + \rho^+ / r^+_{i,o} = 1+x $ varies only slightly between $1$ at the wall and its largest value at mid-gap $1 + d^+ / (2 r^+_{i,o})$. We have discussed this already for the axial velocity profile in Sect.\ III. 
Thus the profile slope according to eq.\ (\ref{profileequationwallunits}) again is that of a logarithmic profile $\propto 1/(1+ K^L \rho^+)$, modulated by a reduction factor, which here is $1/r^3$ instead of only $1/r$ as in the case of the wind profile. Therefore for fixed wall distance the curvature 
effects are much stronger and are much better visible in the angular velocity profile, as compared to the 
wind velocity profile. 
 The physical reason for this significantly stronger 
reduction $\propto 1/r^3 $ of the profile slope of the azimuthal velocity than for the axial velocity with $\propto 1/r$ is that the 
azimuthal motion has to follow the curved, circular 
cylinder surface, while the axial motion is along the straight axis of the cylinder.    
The slope reduction is the stronger, the larger the gap or  
the smaller $\eta$, 
reflecting the stronger curvature effect.  
Also, for TC-devices with the same gap width $d$, the reduction is the larger the smaller the inner cylinder radius is.

\begin{figure}[tb]
    \centering         
\includegraphics[scale=0.48]{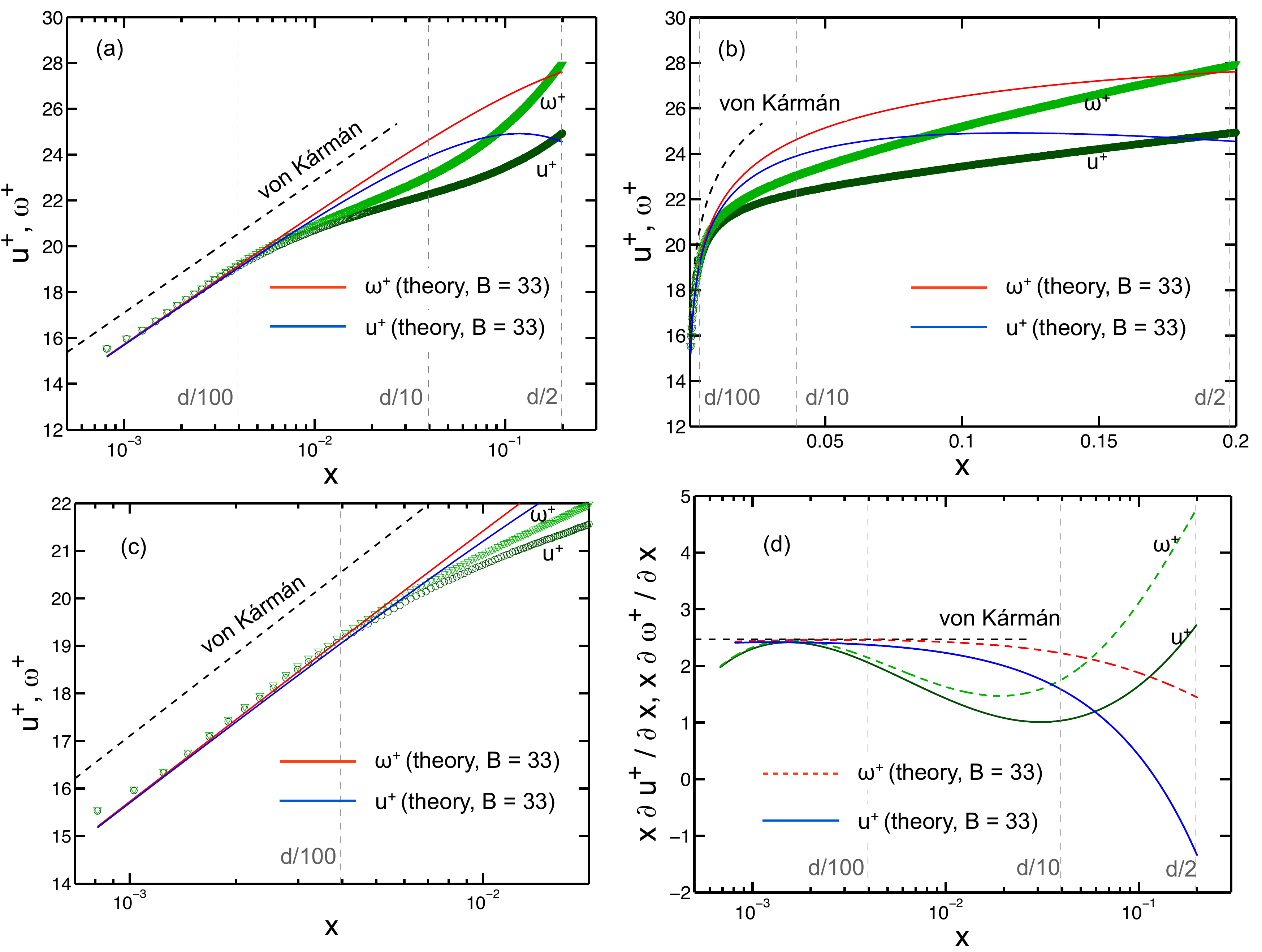}                     
    \caption{(color online)
The universal (i.e., $Ta$-independent) angular velocity profile $\omega^+(x)$ (equation (\ref{reveal-length-scale}))
and the  universal azimuthal  velocity profile $u^+(x)$ (equation (\ref{uprofile-1})) as they  follow from the
present theory on (a) a log-linear scale and (b) a linear-linear scale, in comparison with the experimental data 
from \cite{hui13} for $Ta= 6.2 \cdot 10^{12}$. Note that the representation of figure \ref{fig1}b does {\em not} lead to universal curves.
(c) Zoom-in of (a). 
In (d) the corresponding compensated plots 
$x du^+/dx$ and 
$x d\omega ^+/dx$ are given.  Note that $x d\omega^+ /  d x = d \omega^+/ d (ln x)$. 
}
\label{fig3}
\end{figure}

To analytically calculate the angular velocity profile in detail, one has to integrate the profile equation 
(\ref{profileequationwallunits-prime}). This can be done analytically by 
employing decomposition into partial fractions. 
We will do so using 
the fact that in general $1/r_i^+ \ll K^L x$ or $1 \ll K^L \rho^+$. Then the partial fraction decomposition of the rhs of eq.\ (\ref{profileequationwallunits-prime}) reads (apart from the factor $1/K^L$) 
\be
\frac{1}{x(1+x)^3} = \frac{A}{x} + \frac{B_3}{(1+x)^3} + \frac{B_2}{(1+x)^2} + \frac{B_1}{1+x}. \label{decom}
\ee
The coefficients can be calculated by multiplying with the denominator on the lhs, leading to 
\be
1 = A (1+x)^3 + x \left[ B_3 + B_2 (1+x) + B_1(1+x)^2  \right] .
\ee
The four coefficients can all be calculated by comparing the respective $x$-powers. The result is $A = 1, B_i = -1$ for 
all $i = 1,2,3$. They of course do not depend on any system parameter. 
We now integrate equation (\ref{profileequationwallunits-prime}) for the slope of the angular velocity 
with the decomposition (\ref{decom}) term by term and get 
\be
\omega^+ (x) = {1\over K_L} 
\left(
~\mbox{ln}~x + \frac{1/2}{(1+x)^2} + \frac{1}{(1+x)} - ~\mbox{ln}(1+x)  \right) + B^\prime. 
\label{reveal-length-scale}
\ee
The last term $B^\prime$ is part of the usual additive shift in the log-regime 
 and is determined from experiment. However, also the 2nd and 3rd term contain such an additive
 shift, namely $1.5/K_L$, which we absorb in $B^\prime$ (which then we call $B$) to finally obtain the main 
result of this paper, namely 
the universal (i.e., $Ta$-independent) angular 
velocity profile 
 \be
\omega^+ (x) = {1\over K_L} 
\left(
~\mbox{ln}~x - \frac{x(2+3x)}{(1+x)^2} -  ~\mbox{ln}(1+x)  \right) + B. 
\label{reveal-length-scale-prime}
\ee
The angular velocity profile thus is a log-profile with downward corrections; it is plotted
 in figure \ref{fig3} in various representations. 
For very small $x$ the first log-term will dominate. This slowly increasing log-term will -- with increasing $x$ -- 
be turned downwards, representing the downward trend of the profile.  
In the limit $r^+_{i,o} \rightarrow \infty$, channel flow, we have 
$x  \rightarrow 0$ and the mere log-profile $\propto (K_L^{-1} \ln x + B) $ is recovered. 
Equation (\ref{reveal-length-scale-prime}), together with the definition (\ref{x}) of the dimensionless length $x$, 
 thus nicely reveals that in general there is an  extra intrinsic lengthscale $r^+_{i,o}$ 
 in the TC profile,  in contrast to that for  plate flow.

From eq.\ (\ref{logical-omega-plot}) we can now also calculate the universal (i.e., $Ta$-independent) 
azimuthal velocity profile, 
\be \label{uprofile-1}
u^+ (x) =  \left( 1 + x  \right) \omega^+   -  F_i^{-1}    x  .     
\ee
which is also shown in figure \ref{fig3} in various representations. This relation between $u^+$ and $\omega^+$ has already been given in another form in eq. (\ref{logical-omega-plot}). 
Due to the extra factor $1 + x $ in front of $\omega^+$ and due to the additive term, {\it not both} 
profiles,  $\omega^+ (x) $ {\it and} $u^+ (x)$, can be log-laws. 
Since within our framework $\omega^+(x) $ {\it is} a log-law, see eq.\ 
(\ref{reveal-length-scale-prime}), thus 
$u^+(x) $ can{\it not}  be. 
For smaller gaps ($\eta$ not too far from 1) this will not be visible 
because of the experimental scatter and finite precision, 
but for larger gap (thus smaller $\eta$), the azimuthal velocity profile $u^+ (x) $ 
 will clearly deviate from the log-law of the wall. 

The difference between the $\omega^+$-profile and the $u^+$ profile can nicely be seen from figures \ref{fig3}a,c,d:
The curve for $\omega^+$ follows the ideal von K\'arm\'an log-law (straight line in figs.\ \ref{fig3}a,c and straight horizontal line in fig.\ \ref{fig3}d) much longer than that one for $u^+$, before the curvature corrections for large $x$ set in and 
bend down the $\omega^+ (x)$ curve -- for the $u^+$-curve the deviations from the log-law set in earlier and are stronger.

\section{Comparison with experimental data}

\begin{figure}[tb]
    \centering         
 \includegraphics[scale=0.5]{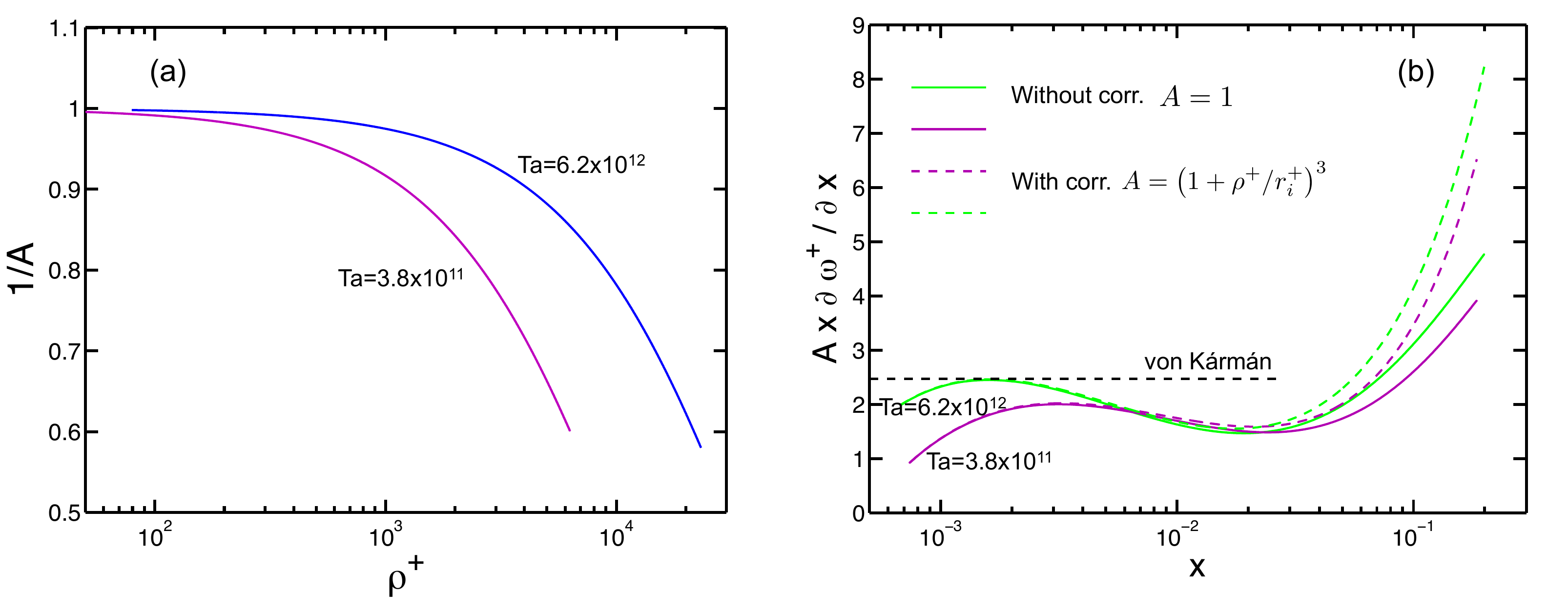}                     
    \caption{(color online)
(a) 
(Inverse) correction factor $A = \left(1 + {\rho^+ \over r_i^+} \right)^3$
as function of $\log_{10} \rho^+$ for the two Taylor numbers of figure \ref{fig1}a; here $\eta = 0.7158$ as in ref.\ \cite{hui13}. 
%!!!!!!!!!WE MAY WANT TO INCLUDE ETA = 0.5 HERE, TOO, TO SHOW HOW STRONG THE EFFECT DEPENDS ON THE CURVATURE. DL. !!!!!!!!!!!!!!
The correction factor $A$ is applied to the angular velocity and in (b) we show the compensated angular velocity slope 
$x d\omega^+/dx $ vs $x=\rho^+/r_i^+$ on a log-linear scale. 
}
\label{fig2}
\end{figure}

To further quantitatively illustrate the results in a better way, we use
 the geometrical parameters of the
 $T^3C$ facility \cite{gil11a}, which has $\eta = 0.7158$. 
The curvature correction factor  
\be 
A = 
\left(1 + {\rho^+ \over r_i^+} \right)^3 
= \left(1 + {\rho \over r_i} \right)^3 
= (1+x)^3
\label{correction-factor}
\ee
for the angular velocity slope (eq.\ (\ref{profileequationwallunits})) 
is shown in figure \ref{fig2}a for two different $Ta$ of the experiments of ref.\ \cite{hui13}.
At the end of the log-range 
(assumed to be at $d/100$)
we  find an angular velocity slope decrease 
by a factor of $0.9882$, which would clearly be hard to visualize. 
For small gap samples, say with $\eta = 0.9$, the correction factor will be even closer to one.
In contrast, 
if $\eta = 0.5$, which numerically is available with DNS, at the end of the log range (again assumed to be d/100)
the log-slope is reduced by a factor of 0.9706, which may become visible. 
For larger wall distances $\rho \gg d/100$ 
the correction factor $A$ gets visibly smaller than 1.  
However, for these distances it was found experimentally
that one is already far away from a log-range, see figure \ref{fig1}. 
In figure \ref{fig2}b we apply the correction factor to the angular velocity slope 
(\ref{profileequationwallunits-prime}), which is universal (i.e., independent of $Ta$) for large wall distances. 
We see that for small wall distances 
 the correction factor 
indeed brings the compensated profile closer to the log-profile (i.e., a horizontal line in this plot), but that 
this curvature effect is very small. The correction factor only 
becomes substantial close to the outer scale $\rho \sim d/2$ where the log-regime clearly has already ceased.

In figure \ref{fig3}, in addition to the universal theoretical profiles for $\omega^+(x)$ and $u^+(x)$, 
we also include the experimentally  measured \cite{hui13} profiles for the largest available Taylor number
$Ta = 6.2 \cdot 10^{12}$. We see that the experimental curves {\it qualitatively} follow the same trend as the theoretical ones.
In particular, the $\omega^+(x)$ profiles are closer to the log-profiles as the $u^+(x)$ profiles, and 
both show increasing deviations from the log-profiles for increasing $x$. 
However, there are pronounced  {\it quantitative} differences between theory and experiment: First of all, for very large $x\sim 0.1$
the experimental profiles bend up again. This is to be expected as then one is already very close to the gap center
and the effect of the opposite side of the gap becomes relevant -- one is then simply far away from the boundary layers.
But second, and more seriously, already at $x\sim 4 \cdot 10^{-3}$, corresponding to a wall distance of $d/100$, 
the quantitative deviations between theory and experiment become very visible.

What are the reasons for the quantitative descrepancies between theory and experiments? 
The theory has made certain assumptions like the existence of a turbulent $\omega$-diffusivity and its functional dependence 
(\ref{omega-diffusivity}) 
on the wall distance. We consider this as a relative innocent assumption. More seriously is the fact that the
theory does not take full notice of the Taylor roll-structure of the flow and the resulting flow inhomogeneity in vertical 
(z-)direction. From the numerical simulations of Ostilla {\it et al.}\ \cite{ost14} we know however that the boundary layer profiles
pronouncedly depend on the vertical direction, at least up to Taylor numbers $Ta \sim 10^{10}$ (larger ones are presently numerically
not yet achievable), but presumably beyond. Log-layers develop in particular in the regions in which plumes are emitted and in shear
layers, but not in regions in which plumes impact. For increasing $Ta$ the height dependence gets weaker, but 
it still persists at
 $Ta = 6.2 \cdot 10^{12}$ \cite{hui12}, for which we present the experimental data here and which is the largest available Taylor number.
In fact, 
as seen from figure 2b of ref.\ \cite{hui12}  
at $Ta = 1.5 \cdot 10^{12}$ the local Nusselt number can vary from height to height even up to a factor of two,
with the corresponding consequences on the local profiles. 
The reason for the persistence of the vertical dependence is the limited mobility of the Taylor rolls, due to their confinement 
between the upper and lower plates. Though the aspect ratio -- TC cell height dived by gap width $d$ -- 
in experiment is $\Gamma = 11.7$ \cite{hui12}, the six to eight Taylor rolls are still relatively fixed in space. The 
profile measurements of ref.\ \cite{hui13} took place at fixed position, namely mid-height. The theoretical results should be
understood as some height-averaged results, and strictly speaking such height-averaging only makes sense if there is no or hardly 
any vertical dependence. 

Clearly, it would be of utmost importance to {\it measure} the height dependence of the angular velocity and azimuthal velocity profiles,
in order to quantify it and 
to see whether the relatively poor quantitative agreement between theory and experiment is better at other heights. 
Also numerical simulations to further explore the height dependence of the profiles would be useful, similar to 
what had already been done in ref.\ \cite{ost14}, but now for both $\omega^+(x)$- and $u^+(x)$-profiles and also
profiles of the vertical (i.e., wind) velocity, for even larger $Ta$, for 
smaller $\eta$, and finally for different values of co- and counter-rotation, i.e., different $\mu$, but of course all
in the turbulent regime. 
Work in this direction is on its way.

Finally, we estimate
 the wall parameters, the amplitude $u^*$ of the azimuthal velocity fluctuations, and the slope parameter $F_i$, all based on experimental data.
 In the experiment \cite{hui13} the outer cylinder was kept at rest, $\omega_o = Re_o = 0$. Then (at the inner cylinder) $(u^*)^2 = N^{\omega} J^{\omega}_{lam} / r_i^2 = N^{\omega} ~\nu r_o^2 \omega_i / (r_a \cdot d)$. 
Expressed in terms of $Ta$ (for $\omega_o = 0$) one has $\omega_i = \nu r_g^2 \sqrt{Ta} / (r_a^3 d)$ and $Re_i = (\eta^2 / (\frac{1 + \eta}{2})^3) \sqrt{Ta} = 0.8115 \sqrt{Ta}$; here $r_a, r_g$ are the arithmetic and geometric mean radii. The Nusselt number $N^{\omega}$ as function of the Taylor number $Ta$ for the $T^3C$ apparatus has already been measured for $T^3C$ in \cite{gil12}. Putting all together leads to 
\be \label{innerfluctuationvelocity1}
u^*_i = \sqrt{6.81\times10^{-3}} ~\frac{\nu r_o r_g}{r_a^2 d}~Ta^{0.4375} ~.
\ee 
Inserting the material and geometry parameters of $T^3C$ as given above one obtains an explicit expression for $u^*_i$ ,
\be \label{innerfluctuationvelocity2}
u^*_i = 1.1948 \times 10^{-6} \times ~Ta^{0.4375} ~\mbox{ms}^{-1} ~.
\ee
This allows us to determine all physical parameters of interest. They are compiled in table \ref{table2}. 
%!!!!!!!!!!!!!TABLE MUST BE DISCUSSED MORE. !!!!!!!!!!!

 \begin{table}[tb]
 \begin{center}
 \begin{tabular}{|c|c|c|c|c|c|c|c|c|}
 \hline
          $Ta$
       &  $u_{i}^*$ in ms$^{-1}$ 
       &  $Re^*_{\omega,i} = \frac{{u^*_i} \cdot d}{\nu}$ 
       &  $\delta_{i}^*=\frac{\nu}{u_{i}^*}$ 
       &  $\frac{r_i}{\delta^*_{i}}$
       &  $\frac{d/2}{\delta^*_{i}}$
       &  $\frac{d/100}{\delta^*_{i}}$
       &  $F_i = \frac{Re^*_{i}}{Re_i} = \frac{\omega^*_i}{\omega_i}$
       &  $\frac{u^*_z}{u^*_i}$
\\
\hline
%          $4.6 \times 10^7$
%       &  ~~0.0027
%       &  ~~~213
%       &  $370 \times 10^{-6}$m
%       &  ~~~~541
%       &  ~~~107
%       &  ~~~2
%       &  $3.87 \times 10^{-2}$
%       &  1.33
%\\
%             $1 \times 10^9$ 
%       &  ~~0.0104
%       &  ~~~821 
%       &  $96.2 \times 10^{-6}$m
%       &  ~~2 080   
%       &  ~~~411
%       &  ~~~8
%       &  $3.20 \times 10^{-2}$
%       &  1.38
%\\
           $6 \times 10^{10}$
       &   ~~0.0620 
       &   ~4 900
       &    $16.2 \times 10^{-6}$m
       &   12 300  
       &   ~2 440
       &   ~49
       &   $2.47 \times 10^{-2}$
       &   1.48  
\\    
           $4.6 \times 10^{11}$
       &   0.151
       &   12 000
       &   ~~$6.61 \times 10^{-6}$m
       &   30 300
       &    ~5 980
       &    ~120
       &   $2.18 \times 10^{-2}$
       &   1.54
\\
           $3 \times 10^{12}$
       &   0.344 
       &   27 200
       &   ~~$2.91 \times 10^{-6}$m
       &   68 700
       &   13 540
       &   271
       &   $1.94 \times 10^{-2}$
       &   1.61
\\
          $6 \times 10^{12}$  
       &  0.465   
       &  36 700 
       &  ~~$2.15 \times 10^{-6}$m
       &  93 100
	   &  18 420
	   &  368
	   &  $1.85 \times 10^{-2}$
	   &  1.65
\\
\hline
 \end{tabular}
 \end{center}
\caption{ The longitudinal velocity fluctuation amplitudes $u^*_i$ (second column) in the inner cylinder boundary layer for some Taylor numbers $Ta$ (first column). Third column the respective values of the fluctuation Reynolds number $Re^*_{\omega,i}$ in the inner cylinder BL. The fourth column shows the respective viscous length scales $\delta^*_{i} = \nu / u^*_i$~. The fifth column offers the inner cylinder radius 0.2000 m in the respective $\delta^*_{i}$-wall units, known as $r_i^+$. The gap half width in these wall units (sixth column) is $(d^+/2 =) \frac{d/2}{\delta^*_{i}} = \frac{d}{2r_i} r_i^+ = \frac{1}{2}(\eta^{-1} -1) r_i^+ = 0,1985 r_i^+$, i.e., $d^+/2$ is about $r_i^+/5$. Column seven shows $d^+/100$, below which the experimental data
are closest to a log-law. 
Column eight shows $F_i$, which equals $Re^*_{\omega,i}/Re_i = \omega^*_i / \omega_i$, the relative $\omega$-fluctuation amplitude. The last 
(nineth) 
column compiles the factor $u^*_z / u^*_i$, the ratio of the transversal and longitudinal velocity fluctuation amplitudes; this is obtained with the values from column 2 of this table and column 4 of Table \ref{table1}.}      
\label{table2}
\end{table}

%For the readers' convenience we give  the 
%relative magnitudes of $F_i^{-1}$ in equation (\ref{uprofile-1}) as a function of $Ta$ in Table III.
%IS THIS REALLY NECESSARY - ALSO THE TABLE? - I SUGGEST TO REMOVE THE TALBE. DL.
%
% \begin{table}[h]
% \begin{center}
% \begin{tabular}{c|c|c|c|c|c|c}
%% \hline
%          $Ta$
%       &  $4.6 \times 10^7$
%       &  $1 \times 10^9$ 
%       &  $6 \times 10^{10}$ 
%       &  $4.6 \times 10^{11}$
%       &  $3 \times 10^{12}$
%       &  $6 \times 10^{12}$
%\\
%\hline
%          $1 / F_i$
%       &  25.84
%       &  31.25
%       &  40.49
%       &  45.87
%       &  51.55
%       &  53.76
%
%%\\
%%\hline
% \end{tabular}
% \end{center}
%\caption{Depending on $Ta$ the azimuthal velocity $u^+$ experimentally measured in ref.\cite{hui13} has an additional shift of 
%$x=\frac{\rho^+}{r^+_i}$ times a factor $F_i^{-1}$ roughly between 26 to 54. 
%!!!!!!!!!!!!  I SUGGEST TO REMOVE THE TABLE. DL. !!!!!!!!!!!!!!
%}       
%\label{table3}
%\end{table}

\section{Concluding remarks}
In summary, we have derived a Navier-Stokes-based theory for the velocity and angular velocity profiles in turbulent TC flow,
following the same approach as that one of ref.\ \cite{gro12} for RB flow, but in cylinder geometry appropriate for TC flow, taking proper care of the wall curvature(s). 
The main findings are  
\begin{itemize}
\item that the angular velocity profile follows an universal  log-law (eq.\ (\ref{reveal-length-scale-prime}), reflecting the 
curvature corrections), 
\item that the universal azimuthal velocity profile eq.\ (\ref{uprofile-1}) 
correspondingly can{\bf not} follow a log-law,
\item and that also the axial velocity profile follows an universal log-law eq.\ (\ref{windprofile}), but with weaker curvature corresions, due to the 
less pronounced effect of the curvature in the flow direction. 
\end{itemize}
Though the experimentally measured angular velocity and azimuthal velocity 
profiles at fixed mid-height {\it qualitatively} follow above trends, the {\it quantitative} agreement is not particularly good as the measured
deviations from
the log-law are much stronger. This could be due to the roll structure of the flow, leading to height dependences of the flow
profiles, which are not considered in the theory. We finally suggest various further experimental and numerical measurements 
to further validate for falsify the presented theory.

\section*{Acknowledgement}
We thank Rodolfo Ostilla and Sander Huisman for various insightful 
discussions on the subject and all authors of ref.\ \cite{hui13} for making
the data of that paper available for the present one. 
We finally acknowledge FOM for continuous support of our turbulence research.

\begin{appendix}
\section{Alternative choice of transversal fluctuation amplitude}
We here check the idea that the relevant velocity for determining the transversal fluctuation amplitude $u^*_{w,z}$ is {\it not} the inner cylinder rotation velocity $U_i$ but, instead, the coherent flow -- or wind -- due to the remnants of the rolls, called $U_w$. Its amplitude has been given for the $T^3C$ facility in \cite{gil12}, page 130, to be $Re_w = 0.0424 \times Ta^{0.495}$; this holds for $a=0$ and in the range $3.8 \times 10^9 \lesssim Ta \lesssim 6.2 \times 10^{12}$. This rather small value for the wind Reynolds number rests on PIV measurements, cf.\  \cite{hui12}. A similar but even smaller value has been obtained with DNS, \cite{ost13}: In the regime $4 \times 10^4 \lesssim Ta \lesssim 1 \times 10^7$ it is $Re_w = 0.0158 \times Ta^{0.53}$. -- The relevant quantities are given in Table \ref{table4}.

 \begin{table}[tb]
 \begin{center}
 \begin{tabular}{|c|c|c|c|c|c|c|}
 \hline
          $Ta$
       &  $Re_w$ 
       &  $U_w = Re_w \frac{\nu}{d}$ 
       &  $\frac{U_w}{U_i}$ 
       &  $\frac{u^*_{w,z}}{U_w} = \frac{\bar{\kappa}}{W(3.2 Re_w)}$
       &  $u^*_{w,z} = U_w \frac{\bar{\kappa}}{W(3.2 Re_w)}$
       &  $\frac{u^*_{w,z}}{u^*_i}$
\\
\hline  
%          $10^8$ 
%       &  387 
%       &  0.0049 m/s
%       &  0.0476
%       &  0.07367 
%       &  0.3610 $\times 10^{-3}$ ~m/s
%       &  -- 
%\\
%          $10^9$
%       &  1209 
%       &  0.0153 m/s
%       &  0.0471 
%       &  0.06246
%       &  0.9556 $\times 10^{-3}$ ~m/s
%       &  --   
%\\
%          $10^{10}$  
%       &  3779 
%       &  0.0478 m/s 
%	   &  0.0465
%	   &  0.05406
%	   &  2.5841 $\times 10^{-3}$ ~m/s
%	   &  --
%\\
%          $10^{11}$
%       &  11813
%       &  0.1495 m/s
%       &  0.0460
%       &  0.04756 
%       &  7.1102 $\times 10^{-3}$ ~m/s
%       &  -- 
%\\    
%          $10^{12}$	
%       &  36929
%       &  0.4675 m/s
%       &  0.0455
%       &  0.04239
%       &  19.8173 $\times 10^{-3}$ ~m/s
%       &  --   
%\\
%\hline
          $6 \times 10^{10}$
       &  9174
       &  0.1161 m/s
       &  0.0461
       &  0.04887
       &  5.6738 $\times 10^{-3}$ ~m/s
       &  0.092
\\
          $4.6 \times 10^{11}$
       &  25144
       &  0.3183 m/s
       &  0.0456
       &  0.04401
       &  14.0084 $\times 10^{-3}$ ~m/s
       &  0.093
\\     
          $3 \times 10^{12}$
       &  63612
       &  0.8052 m/s
       &  0.0452
       &  0.04029
       &  32.4415 $\times 10^{-3}$ ~m/s
       &  0.094
\\
          $6 \times 10^{12}$
       &  89650
       &  1.1291 m/s
       &  0.0451
       &  0.03906
       &  44.1026 $\times 10^{-3}$ ~m/s
       &  0.095
\\            
\hline

 \end{tabular}
 \end{center}
 \caption{The wind fluctuation amplitude $u^*_{w,z}$ based on the wind Reynolds number for different Taylor numbers (column 1). The corresponding wind Reynolds number $Re_w = 0.0424 \cdot Ta^{0.495}$ are listed in column 2. The next columns, 3 and 4, show the respective wind velocities $U_w = Re_w \cdot \nu / d $ in m/s (with system parameters $\nu = 1 \cdot 10^{-6}$ m$^2$s$^{-1}$ and $d=0.0794$ m) as well as the wind velocity relative to the inner cylinder rotation velocity, $ U_w /U_i $, turning out to be of order 4.6\%, only very slightly depending on $Ta$; it is $U_w / U_i = 0.052193 \cdot Ta^{-0.005}$, were we used $Re_i = 0.8115 ~Ta^{0.5}$ for $a=0$, as was reported in the main text. Column 5 offers the relative fluctuation amplitude $ u^*_{w,z} /U_w $, obtained from the formula with Lambert's W-function derived in \cite{gro11}, 
$u^* _{w,z}/ U_w = \bar{\kappa} / W(3.2 Re_w)$ (again we have used  $\bar{\kappa} = 0.4$ as von K\'arm\'an's constant and $b = 0.125$ to determine the argument $\frac{\bar{\kappa}}{b}Re_w$ of $W$. Column 6 contains the respective transversal fluctuation amplitudes $u^*_{w,z} = U_w \cdot \frac{\bar{\kappa}}{W(3.2 Re_w)}$. Finally in column 7 the ratios of the transversal wind fluctuation amplitude 
$u_{w,z}^*$
and the longitudinal angular momentum fluctuation amplitude 
$u_{i}^*$
are offered for the $Ta$-values of interest from Table \ref{table2}. Note that the transversal wind fluctuation amplitude is about 9.5\% of the longitudinal $\omega$-fluctuation amplitude, determined in Section IV, pretty independent of $Ta$.} 
\label{table4}
 \end{table}

\end{appendix}

%\bibliographystyle{prsty_withtitle}
%\bibliography{literatur}

\end{document}